\documentclass[aps,pra,twocolumn,groupedaddress, showpacs]{revtex4-1}
\usepackage{graphics,graphicx,float,amsmath, xcolor} 
\usepackage{upgreek}
\usepackage{hyperref}

\newcommand{\beq}{\begin{equation}}
\newcommand{\eeq}{\end{equation}}
\newcommand{\um}{$\upmu$m}

\begin{document}


\title{A Fabry-Perot Microcavity for Diamond-Based Photonics}


\author{Erika Janitz}
\author{Maximilian Ruf}
\author{Mark Dimock}
\author{Alexandre Bourassa}
\author{Jack Sankey}
\author{Lilian Childress}
\affiliation{Department of Physics, McGill University, Montreal CA}


\date{\today}

\begin{abstract}
Open Fabry-Perot microcavities represent a promising route for achieving a quantum electrodynamics (cavity-QED) platform with diamond-based emitters. In particular, they offer the opportunity to introduce high purity, minimally fabricated material into a tunable, high quality factor optical resonator. Here, we demonstrate a fiber-based microcavity incorporating a thick ($>10$ \um)~diamond membrane with a finesse of 17,000, corresponding to a quality factor $Q\sim10^6$. Such minimally fabricated, thick samples can contain optically stable emitters similar to those found in bulk diamond. We observe modified microcavity spectra in the presence of the membrane, and develop analytic and numerical models to describe the effect of the membrane on cavity modes, including loss and coupling to higher-order transverse modes. We estimate that a Purcell enhancement of approximately 20 should be possible for emitters within the diamond in this device, and provide evidence that better diamond surface treatments and mirror coatings could increase this value to 200 in a realistic system.  \end{abstract}

\pacs{42.81.Wg, 42.25.Hz, 42.60.Da, 42.50.-p}

\maketitle

\section{Introduction}


The field of diamond photonics has seen tremendous growth over the last decade~\cite{Loncar2013, Aharonovich2014}, spurred by new applications of optically active defect centers in metrology~\cite{Degen2014, Rondin2014} and quantum information science~\cite{Ladd2010, Nemoto2014}.  In particular, the nitrogen-vacancy (NV) defect~\cite{Doherty2013} exhibits long spin coherence times and narrow optical transitions favorable for realizing a solid-state cavity-QED system. In pursuit of this goal, much progress has been made in fabricating low-mode-volume cavities in diamond itself~\cite{Aharonovich2014, Burek2014, Khanaliloo2015, Lee2014}, and Purcell enhancement of the NV zero phonon line as large as 70 has been observed in a diamond photonic crystal cavity~\cite{Faraon2012}. A complementary strategy is to confine the defect in an open Fabry-Perot microcavity~\cite{Hunger2010_NJP, Muller2010}, which provides in-situ tunability and the possibility for very narrow cavity linewidths. Recently, three groups have observed coupling between an open cavity and an NV center in a nanocrystal~\cite{Kaupp2013, Albrecht2013, Albrecht2014, Johnson2015}. 

A central challenge for diamond photonics is destabilization of defect optical transitions in close proximity to surfaces, especially for defects in nanocrystals 
or in nanofabricated devices~\cite{Santori2010}. For example, the aforementioned diamond photonic crystal cavity achieved its high Purcell factor at the expense of spectral diffusion of many GHz~\cite{Faraon2012}, far in excess of the near lifetime-limited linewidths of ~13 MHz that can be observed in some type IIa samples~\cite{Tamarat2006}.  While recent advances in surface treatments~\cite{Chu2014} and fabrication~\cite{Mouradian2015} hold promise for realizing optimal NV properties in nanophotonic structures, narrow linewidths are most reliably obtained microns into bulk diamond. 
The open cavities discussed here can accommodate the larger mode volumes required for such microns-thick samples: both their mode volume $V$ and quality factor $Q$ increase approximately linearly with length, so that the Purcell enhancement $F_p$ depends only on the mirror finesse $\mathcal{F}$ and the ratio of the cavity waist $w_0$ to the resonant wavelength $\lambda$: $F_p\propto \mathcal{F} \lambda^2/w_0^2$~\cite{Hunger2010_NJP}. In addition, their linewidths are comparatively narrow and can be tuned over a wide range in-situ via the cavity length, potentially allowing exploration of spin-dependent coupling between an NV center and the cavity. Finally, by adjusting the positions of the mirrors, one can optimize the cavity mode spatial overlap with the emitter. 

In principle, it is straightforward to incorporate a microns-thick membrane into an open cavity.  However,  for high finesse $\mathcal{F}>10^4$ cavities, losses at the 100 ppm level are important. Absorption and scattering must be minimized, and changes in the cavity mode induced by the dielectric interface must be considered. Here, we demonstrate that a fiber-based microcavity can maintain high finesse $\mathcal{F} \sim 17,000$ (quality factor $Q\sim 10^6$) while incorporating a $>10$ \um~thick diamond membrane compatible with high stability defect centers. We further develop a theoretical description for the longitudinal modes (including diffraction effects), and perturbatively estimate the mixing between transverse modes induced by the membrane. Based on our measurements, we predict our device should be capable of enhancing the NV zero phonon line by a factor of approximately 20.

\section{The Fiber Cavity Device}
\label{device}
\begin{figure}[h!]
\includegraphics[scale=1]{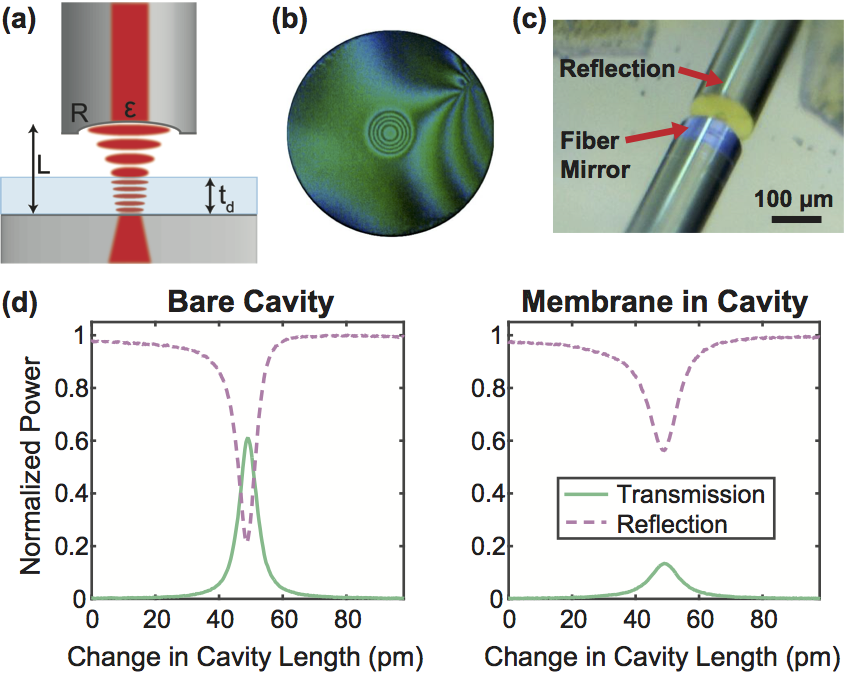}
\caption{(Color online) (a) A schematic of the microcavity system illustrating the cavity length ($L$), mirror radius of curvature ($R$), diamond membrane thickness ($t_d$), and coupling efficiency ($\epsilon$) between the fiber and cavity modes. (b) An interferometric image of an ablated fiber, where each subsequent dark fringe corresponds to a change in depth of $\approx 253$ nm. 
(c) A microscope camera image of the fiber cavity mirror and reflection seen against the diamond membrane. Dark regions in the upper corners are due to partial etching of the mirror beneath the membrane, and some contamination of the diamond surface is also visible. (d)  Measured transmission (solid line) and reflection (dashed line) curves for the bare cavity, and membrane-in-cavity configurations, normalized to the peak reflected power. These particular measurements correspond to finesses of  $\mathcal{F}_L\approx 37,000 \pm1000$ and $\mathcal{F}_L\approx 15,900 \pm 900$ for the bare cavity and membrane-in-cavity respectively. \label{fig1}}
\end{figure}

We work with a fiber-based Fabry-Perot microcavity~\cite{Hunger2010_NJP, Muller2010} in a geometry similar to those used to study quantum dots~\cite{Miguel2013} and molecules~\cite{Toninelli2010}. The microcavity system (Fig.~\ref{fig1}a) consists of a 
concave mirror on the tip of a single mode optical fiber, and a macroscopic flat mirror to which we bond the diamond membrane. Compared with traditional optics, these fiber-based cavities offer advantages in stable alignment and efficient coupling 
to the single mode propagating in the fiber~\cite{Hunger2010_NJP}. 

The fiber mirror substrate is fabricated using a CO$_2$ laser ablation process \cite{Hunger2010_NJP}. The ablation laser induces evaporation and melting of the glass on the fiber tip \cite{Hunger2012_Advances}, resulting in an approximately Gaussian-shaped dimple (Fig.~\ref{fig1}b) with an extremely low surface roughness of $<0.2$ nm-rms, as measured with an atomic force microscope. By imaging the fiber core during ablation alignment, we achieve a placement repeatability of 0.5 \um~for the ablation spot.  The fiber used in our experiments is measured to have a power coupling efficiency to the cavity mode of $\epsilon^2=48\pm4\%$ (limited by mirror absorption as well as cavity and ablation misalignment, see Appendix~\ref{appendix_coupling}), and an effective radius of curvature $R=61.0\pm1.4$ \um ~(see Appendix~\ref{app_rcurvature}).


The second mirror substrate is a superpolished macroscopic mirror flat with surface roughness below $0.1$ nm-rms. The flat and fiber mirrors are coated with a dielectric mirror stack (LASEROPTIK) specified to have a transmission of $70\pm 10$ ppm and $<24 $ ppm losses at $\lambda = 637$ nm; the theoretical finesse of the stack design is 53,100.  
The fiber (stripped of all polymer jacketing) and flat mirrors were both annealed at $300^{\circ}$C for five hours under atmospheric conditions to reduce losses in the coatings~\cite{Brandstatter2013}.

The diamond membrane 
is fabricated from a $\langle 100 \rangle$-cut electronic grade single crystal diamond plate. 
The bulk diamond was laser cut laterally, 
producing $20 \pm10 \,$ \um~thick diamond membranes polished to a surface roughness of approximately 5 nm-rms. One of the resulting membranes was 
cleaned in a piranha solution and bonded with Van der Waals forces to a silicon carrier wafer, and approximately 
 $2$ \um~was 
etched from the membrane surface using an ArCl$_2$ inductively coupled plasma reactive ion etching (ICP RIE) recipe \cite{Hausmann2012_Nanoletters,Tao2013_AM,Ovar2012_APL}. The etching process reduced the surface roughness to $< 0.2$ nm-rms (measured over one optical wavelength squared). The membrane was then removed from the carrier, similarly cleaned in piranha, etched on the other side (again removing 2 $\mu$m with ArCl$_2$), and finally bonded to the macroscopic mirror flat. Finally, a third ArCl$_2$ etch was performed to thin the membrane  to approximately $10 $ \um. 

To assemble the cavity, the mirror flat is fixed to a tip-tilt mount, while the fiber is clamped to a 3-axis manual and piezo stage. 
The tip-tilt mount enables angular alignment of the cavity mode, while the 3-axis stage allows for study of different regions of the membrane as well as precision control of the cavity length. Figure~\ref{fig1}c shows an image of the assembled device.

\section{Effect of the Membrane on Cavity Modes}
\label{length_meas}

\subsection{Cavity finesse}

Introducing the membrane into the cavity affects the linewidth of its resonances. To determine the cavity finesse, we scan the position of the fiber mirror while monitoring the cavity's transmission and reflection at a fixed wavelength near 637 nm (provided by a tunable diode laser). In this case, we define the finesse as the ratio of the free spectral range (FSR $\approx \lambda/2$) to the resonance full width half maximum (FWHM) measured as a function of the length of the cavity, and denote it by $\mathcal{F}_L$:
\beq
\mathcal{F}_L  = \frac{\mathrm{FSR~in~length}}{\mathrm{FWHM~in~length}}
\eeq
Note that for the membrane-in-cavity system, this is not necessarily the same as finesse obtained by measuring the resonance spacing and linewidth as a function of laser frequency. 

We measure the 
finesse by first performing a long scan of the cavity length to observe the resonance spacing as a function of the voltage applied to the piezo stage. By scanning the length over about 20 \um~(roughly 60 FSR), we can fit the observed resonances to extract the free spectral range and calibrate the piezo stage nonlinearity.  
Subsequent voltage scans over shorter length ranges ($0.6$ \um) provide high resolution data for extracting the cavity linewidth. At each position of interest, we measure 64 transmission and reflection peak data sets to gather statistics on the cavity linewidth. This procedure is followed for all measurements of $\mathcal{F}_L$ presented in Figs.~1-4. 

We characterize the transmission and reflection curves for the TEM$_{00}$ fundamental mode using the same fiber mirror for an empty or ``bare" cavity, and for a membrane-in-cavity, as illustrated by sample data sets in Fig.~\ref{fig1}d. These measurements are corrected for calibrated losses in the measurement apparatus (outside of the cavity), and are normalized to the peak reflected power. To determine cavity linewidth, we fit the transmission and reflection data sets to Lorentzian and Fano lineshapes respectively; in practice, the reflection signal gives better signal to noise and was used to calculate cavity finesse.  The asymmetric resonances seen in reflection can arise from a slight displacement of the ablation dimple from the fiber core  
(see Appendix \ref{appendix_coupling}). 

As discussed below, the observed finesse varies with the length of the cavity, the transverse location on the membrane, and the frequency of the laser. 
At best, we observe a peak finesse of $\mathcal{F}_L\approx 37,000$ for the bare cavity and  $\mathcal{F}_L\approx 17,000$ for the membrane-in-cavity setup. When we observe different locations on the flat mirror's surface, we find the finesse for the empty cavity typically fluctuates by a few thousand, most likely due to surface contamination or spatially-varying surface roughness. With the diamond present, however, the finesse fluctuates  by a much larger factor, with no observable cavity resonances in many locations. At first glance one might presume these fluctuations arise from similar physics, i.e. roughness, contamination, or even crystal defects in the diamond itself. However, as discussed below, such large finesse fluctuations are primarily caused by spatial variations in the diamond layer thickness, which affects the cavity mode structure in an important and predictable way.


\subsection{Mode structure}

\begin{figure*}[htb]
\includegraphics{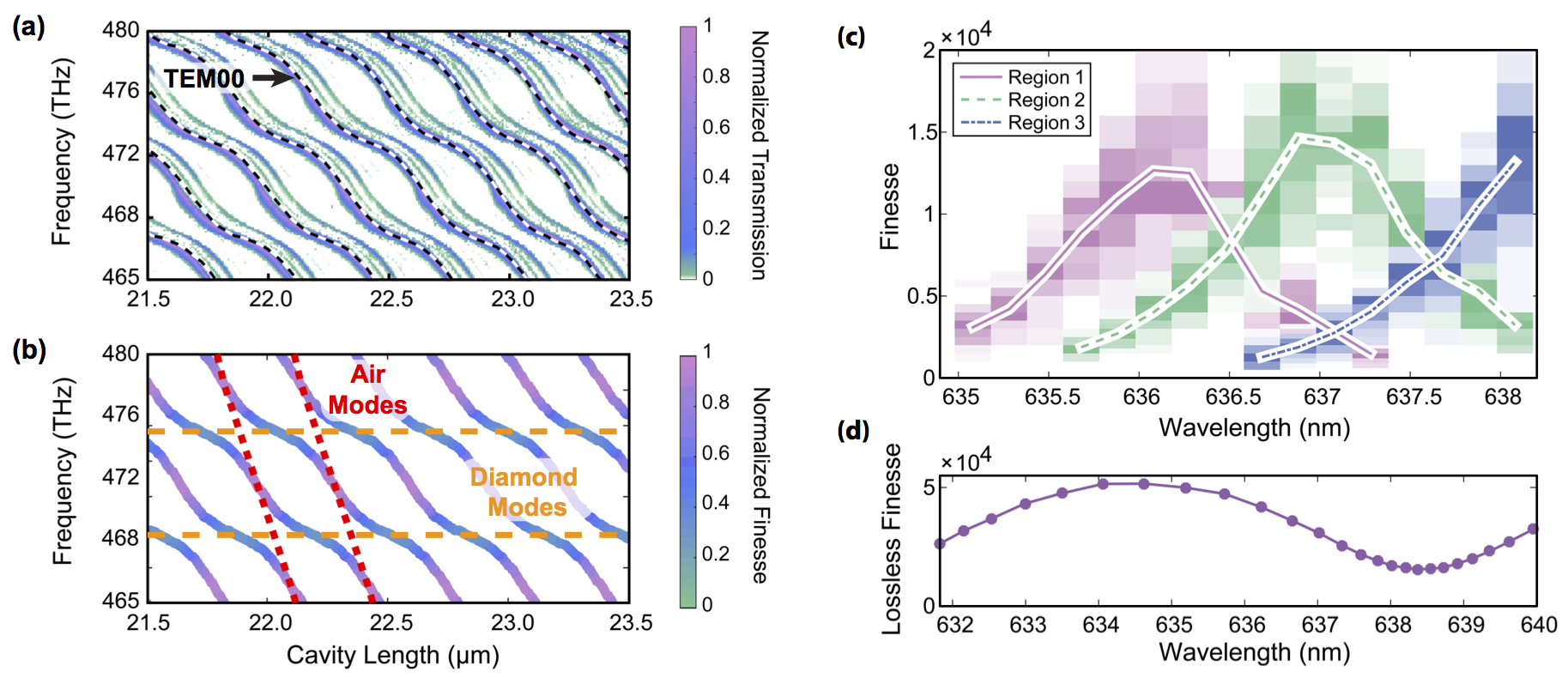}
\caption{(Color online) (a) Cavity spectrum obtained by coupling a broadband LED through the flat mirror, and scanning the length of the cavity. An approximate analytic fit (Eq.~\ref{approxres}) to the resonances is overlaid. The error in the x-axis calibration is $\pm 0.2 \mu$m.  
(b) A numerical simulation of the normalized finesse $\mathcal{F}_L$ of the cavity resonances including the Guoy phase and mirror stack. Lines indicating the cavity resonances associated with air and diamond regions are overlaid to illustrate the avoided crossings. (c) Finesse vs. laser wavelength measured for three different regions of the diamond membrane, corresponding to different membrane thicknesses. The raw finesse data was binned using the Freedman-Diaconis rule to show the underlying distribution, and the binned data is shown through opacity. The mean finesse values are plotted as lines. (d) Simulated finesse vs. laser wavelength for a lossless cavity. The absence of scattering losses leads to high finesse over a much larger wavelength range. \label{fig2}}
\end{figure*}

We characterize the cavity mode structure by illuminating the flat mirror with a broadband LED source,  
and measuring the spectrum of the light transmitted into the fiber with a grating spectrometer. 
By gathering data as a function of cavity length, we observe the evolution of multiple longitudinal and transverse modes (see Fig.~\ref{fig2}a).  

The measured white light spectrum exhibits a canted periodic structure that is markedly different from the behavior of a bare cavity. 
These features can be quantitatively reproduced by a simple one-dimensional (1D) model.  We consider lossless mirrors at each end of the cavity, with a 180 degree phase shift on reflection (facing the cavity) to approximate the dielectric mirror stack terminated at the high index material (Ta$_2$0$_5$ in this case). Between the mirrors are a slab of diamond of thickness $t_d$ and index $n_d$, and a layer of air with thickness $L-t_d$ and index $n_{air} = 1$ (see Fig.~\ref{fig1}a). In the limit of perfect mirrors, the resonant frequencies $\nu$ are given by solutions to the transcendental equation
\begin{eqnarray}
&(1+n_d) \sin{\left(\frac{2\pi \nu}{c}\left(L + t_d (n_d-1)\right)\right)}& \nonumber \\&=(1-n_d) \sin{\left(\frac{2\pi \nu}{c}\left(L - t_d (n_d+1)\right)\right)}&.
\label{resonance}
\end{eqnarray}
Note that while the resonances occur regularly every $c/2\nu$ as the length of the cavity shifts, the variation with frequency is less straightforward.
For long cavities, encompassing many nodes, Eq.~\ref{resonance} can be approximated by writing $\nu$ in terms of its deviation from an integer multiple $m$ of the average free spectral range,  $\nu = \delta \nu + m c/(2(L+(n_d-1)t_d))$, and neglecting $\delta \nu$ in the RHS of Eq.~\ref{resonance}~\cite{Jayich2008}, yielding:
\begin{eqnarray}
&\nu \approx \frac{c}{2\pi \left(L + (n_d-1)t_d\right)} \Bigg\{\pi m \nonumber &\\
&- (-1)^m \arcsin{\left(\frac{n_d -1}{n_d +1}\sin{\left(\frac{m\pi (L - (n_d + 1)t_d)}{L + (n_d-1)t_d}\right)} \right)} \Bigg\}.
\label{approxres}
\end{eqnarray}

Fitting Eq.~\ref{approxres} to the fundamental mode frequencies in the cavity spectrum results in an estimated membrane thickness of $t_d=10.5\ \pm 0.2~\mu$m  and the cavity lengths given in the x-axis of Fig.~\ref{fig2}a, where the fit results are shown by the dashed lines. Note that we fit resonances over the full 20 $\mu$m range of the stage (not shown) to produce these estimates, and included a cubic nonlinearity in the piezo stage response; the region displayed in the figure is representative of the goodness of fit. The fit deviations arise because the model neglects the transverse Gaussian field profile and Guoy phase of the cavity mode, which can lead to errors in the estimated length of up to half a FSR ($\lambda/4$)
\footnote{A significantly more complicated analytic expression including the Guoy phase and transverse mode profile was also derived, but due to the equation complexity and other systematic measurement errors (including stage drift and nonlinearity), the latter model did not improve precision of the membrane thickness and cavity length estimates.}. The fit also allows us to determine the cavity length during transmission and reflection measurements, albeit with an increased uncertainty (roughly $\pm 0.3~\mu$m) \footnote{The increased uncertainty arises because the LED used to obtain the white-light spectrum heats the cavity, and we must allow it to attain thermal equilibrium before acquiring data. We track length changes as the structure warms, but can only calibrate them within $\sim \pm \lambda/4$.}.

The model also provides some intuition about the system. If the membrane-air interface were perfectly reflective (i.e. $n_d \rightarrow \infty$), it would divide the cavity into two, and the normal modes would separate into ``diamond" modes and ``air" modes, wherein the field is entirely localized in either the diamond or air, respectively. Since the diamond thickness is fixed, the diamond mode frequencies  (horizontal lines in Fig.~2b) would not depend on the longitudinal position of the fiber mirror, while the air modes would decrease in frequency as the air gap increases in length (slanted lines in Fig.~2b). Indeed the frequency spacings for these modes would reveal the diamond thickness and cavity length: $\Delta\nu_{di} = \frac{c}{2\,n_d\,t_d}$, and $\Delta\nu_{air} = \frac{c}{2\,(L-t_d)}$ for the diamond and  air modes respectively.  With finite $n_d$, these modes are coupled to one another, leading to the large avoided crossings observed in the spectrum; ``diamond-like" modes have a shallow slope, while the ``air-like" modes have a steeper slope. This behavior is very similar to that of a membrane-in-the-middle system~\cite{Jayich2008}, where the air-diamond dielectric interface plays the role of a weakly reflective, vanishingly thin membrane.


Our analysis above focused on the fundamental mode, and indeed, for an ideal spherical ablation dimple, light from the fiber core should couple primarily to the Gaussian TEM$_{00}$ mode. Nevertheless, some higher order modes are also visible in the spectrum. Similar features observed in the white-light spectrum for the bare cavity are used to extract the effective radius of curvature of the fiber mirror ($R=61.0 \pm 1.4\ $ \um), as noted in section~\ref{device} and detailed in Appendix~\ref{app_rcurvature}. 

In addition to the analytic 1D model used to find the cavity resonance frequencies, we developed a numerical three-dimensional model for the cavity modes that incorporates wavefront curvature within the cavity and the full dielectric mirror stack. This model includes an approximation that the air-diamond interface follows the curvature of the Gaussian wavefronts, in order to prevent coupling between transverse modes via refraction. Figure \ref{fig2}b shows simulated cavity resonances in the absence of any loss, calculated over the same length and frequency range as the white light transmission data. The cavity parameters used are those extracted from the fit in Fig.~\ref{fig2}a, and full calculation details are provided in Appendix \ref{3d_model}. The color of each data point shows the calculated value of $\mathcal{F}_L$ normalized to the naive finesse estimate of $\frac{\pi}{T}$, where $T$ is transmission per pass of one mirror, and all other loss processes are neglected. The highest finesse values are obtained when the laser frequency is tuned to an air-like mode, approaching the naive estimate. Conversely, if the laser frequency is tuned to a diamond-like mode, the measured finesse will be consistently lower than expected. 

We emphasize that our models thus far assume no losses, meaning the aforementioned finesse fluctuations arise entirely from interference effects. The finesse limitations can be understood by considering the effect of  attaching a diamond membrane to the flat mirror. The dielectric coatings used for our mirrors are terminated with a high index material, and are optimized for use in air. Diamond has a high index of refraction ($n_d=2.417$), which effectively lowers the reflectivity of the flat mirror, corresponding to a decrease in finesse for modes in which the electric field is more confined to the membrane. Quantitatively, the lossless 1D analytic model predicts that the finesse of the diamond-like modes is reduced by a factor of $2/(n_d^2 + 1)\approx 0.3$.  Conversely, if low-index-terminated mirrors were used, the diamond-like modes would exhibit the naive finesse while the air-like modes would have finesse reduced by $2/(1/n_d^2 + 1)\approx 0.6$. 

A central prediction of this calculation is that the mode structure can cause dramatic finesse variations with laser wavelength. 
Moreover, because the mode structure shifts with $t_d$, peak finesse values should occur at different frequencies for different membrane thicknesses. Figure \ref{fig2}c shows the measured finesse $\mathcal{F}_L$ as a function of laser wavelength for three regions on the diamond membrane with marginally different thicknesses~\footnote{White light spectra were taken at each membrane region, but the variation in thickness was below our measurement precision of $\sim 200$ nanometers, illustrating that even slight changes in diamond thickness can drastically change the finesse measured for a given laser frequency.}. The square data points correspond to raw finesse data (binned using the Freedman-Diaconis rule to show the underlying distribution) where opacity illustrates measurement frequency. The mean finesse is plotted with a line. For each region, the peak finesse occurs at some wavelength corresponding to an air-like mode. 
The finesse decreases as the laser is tuned away from this wavelength, and the electric field becomes more localized in the diamond membrane. 

While the qualitative features of our data in Fig.~\ref{fig2}c are similar to the lossless model predictions, the drop in finesse is notably larger and steeper. For comparison, Fig.~\ref{fig2}d shows the numerically simulated finesse for the ideal lossless system, which exhibits much more gradual variations. As discussed quantitatively below, the discrepancy can only be explained by including loss primarily at the air-diamond interface, such as scattering from roughness or contamination. The sharp wavelength dependence again arises from interferometric effects: when there is an electric field node at the air-diamond interface, field driven surface losses are strongly suppressed. In this geometry, a node appears at the air-diamond surface only for the air-like mode, providing a mechanism for the sharp finesse peaks in Fig.~\ref{fig2}c.



\begin{figure}[h!]
\includegraphics[scale=1]{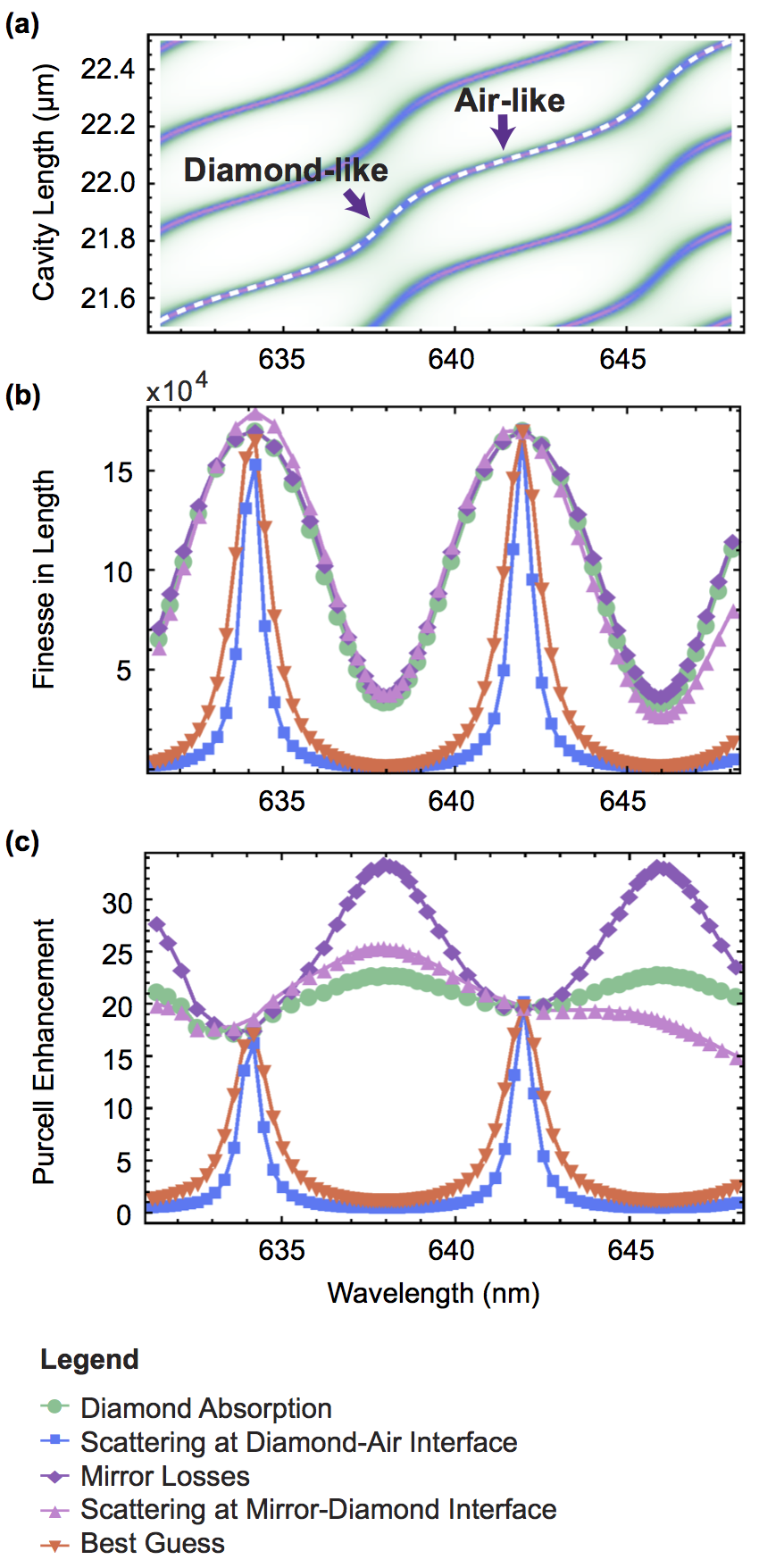}
\caption{(Color online) Numerical cavity simulations including wavefront curvature and loss as a function of resonance wavelength. (a) Simulated cavity transmission, illustrating the cavity lengths for which the simulation was performed (dashed line). Steeper-sloped regions correspond to diamond-like modes, while shallower-slopes correspond to air-like modes.
(b) Simulated finesse $\mathcal{F}_L$ for different sources of loss in the membrane-in-cavity system. For each simulation, enough loss was added to bring the measured finesse in length down to 17,000. (c) Simulated Purcell enhancement for the same models as in (b). Note that the legend at bottom applies to both (b) and (c).  \label{fig3}}
\end{figure}

To quantitatively understand the effects of loss, we add different absorption and scattering mechanisms to the numerical transfer matrix model described in Appendix~\ref{3d_model}. 
We consider loss in the mirrors, loss caused by scattering at the diamond interfaces, and absorption in the diamond.  Loss inside the mirrors and diamond is modeled by adding complex components to the refractive indices of the layers. 
Scattering by surface roughness  of the diamond membrane is added by adjusting the interface reflection and transmission coefficients according to~\cite{SZCZYRBOWSKI1977, Katsidis2002}
\begin{eqnarray}
r_{ij} &=& r^{(0)}_{ij} e^{-2\left(2\pi \sigma n_i/\lambda\right)^2} \label{rij}\\
t_{ij} &=& t^{(0)}_{ij} e^{-(1/2)\left(2\pi \sigma (n_i-n_j)/\lambda\right)^2}, \label{tij}
\end{eqnarray}
where $r_{ij}$ ($t_{ij}$) is the amplitude reflection (transmission) coefficient going from material of index $n_i$ into material of index $n_j$, $\sigma$ is the rms surface roughness, $\lambda$ is the wavelength in vacuum, and $r_{ij}^{(0)}$ and $t_{ij}^{(0)}$ are the lossless Fresnel coefficients. These reflection and transmission coefficients are used in the transfer matrix describing each diamond surface. To quantitatively compare the effects of each individual  source of loss, we increase its strength sufficiently to bring the peak finesse $\mathcal{F}_L$ down to our observed value of 17,000 (while holding other sources of loss at zero). We then calculate the cavity modes and linewidths for the cavity parameters extracted from the fit in Fig.~\ref{fig2}a over the cavity lengths illustrated in Fig.~\ref{fig3}a.

Figure~\ref{fig3}b shows $\mathcal{F}_L$ as a function of wavelength, as predicted by several loss models. As noted earlier, scattering at the air-diamond interface behaves qualitatively differently from the other loss models, and most closely approaches the features we observe in Fig.~\ref{fig2}c. Because there is always a node at the high-index-terminated mirror surface, scattering from the diamond-mirror interface does not produce such sharp features. 

Figures~\ref{fig3}b also includes simulations using our best estimate for the specific losses in our system.  Enough mirror loss was added to bring the finesse down to 37,000 (peak finesse measured for the bare cavity), and we set the diamond-mirror interface roughness to $0.19$ nm-rms (as measured for similar samples). To match the features in Fig.~\ref{fig2}c, we added sufficient scattering at the air-diamond interface to produce a peak in $\mathcal{F_L}$ (see Fig.~\ref{fig3}b) with a FWHM of $1.14$ nm (the linewidth of the central peak in Fig.~\ref{fig2}c). 
Finally, absorptive loss was added to bring the peak $\mathcal{F}_L$ value down to 17,000. Notably, a very large air-diamond surface roughness ($\sigma = 3.5$ nm-rms) was required to reproduce the features of Fig.~\ref{fig2}c. This roughness is far larger than values $< 0.2 $ nm-rms measured on diamond samples etched by ArCl$_2$, and indicates some additional surface scattering or contamination is likely to blame. 

After optical measurements were concluded, AFM measurements of the membrane revealed surface roughness of $\sim1$ nm-rms in the regions of interest (measured over one optical wavelength squared).  The increased roughness was likely caused by the third ArCl$_2$ etch (while bonded to the mirror), which produced noticeable surface damage in some areas of the membrane (see Fig.~\ref{fig1}c); the regions used in these experiments appeared unaffected, but in fact suffered roughening. Surface absorption caused by contamination could also be present, and would behave similarly in our model, exhibiting the same sensitivity to a field node at the air-diamond interface. It is therefore likely that reduced losses could be obtained with better surface preparation. Based on our simulations, state-of-the-art surface roughness $\sigma = 0.19$ nm-rms is compatible with finesse $>10^6$ for air-like modes and $> 50,000$ for diamond-like modes.

\section{Estimation of Purcell Enhancement}
A figure of merit for cavity systems is the Purcell factor $F_p$, which describes the spontaneous emission enhancement into the cavity mode for an optimally placed single emitter. For a cavity with varying refractive index~\cite{Gerard2003, Sauvan2013},
\begin{align}
F_p = \frac{3\,c\,\lambda^2}{4\pi^2\, n_d\, \Delta \nu}\frac{|E_{\textrm{max}}|^2}{\iiint n^2(\mathbf{r})\, E^2(\mathbf{r})\, d\mathbf{r}^3}. \label{purcell}
\end{align}
Here, $\Delta \nu$ is the cavity linewidth in frequency, $\lambda$ is the resonant wavelength, $E(\mathbf{r})$ and $n(\mathbf{r})$ are the electric field and index of refraction within the cavity, and $E_{\textrm{max}}$ is the electric field at the emitter in the diamond, assuming perfect emitter orientation and location. 
Notably, the Purcell factor depends on the linewidth in frequency, not length. 
As we are not currently able to directly measure the cavity spatial mode and $\Delta \nu$, we use our model and measurements of $\mathcal{F}_L$ to provide a theoretical estimate of the Purcell enhancement available in this cavity geometry. 

In our analytic 1D model, it is straightforward to calculate the mode integrals of Eq.~\ref{purcell} and the linewidth $\Delta \nu$ in terms of the mirror finesse $\mathcal{F}$ (a function of only the lossless mirror reflectivity). 
This yields a simple result in the limit of large $\mathcal{F}$:
\begin{eqnarray}
F_p^{(A)} &=& \mathcal{F}\frac{6\lambda^2}{n_d^3 \pi^3 w_0^2} \label{purcell1}\\
F_p^{(D)} &=& \mathcal{F}\frac{12\lambda^2}{(n_d^3 +n_d) \pi^3 w_0^2},
\label{purcell2}
\end{eqnarray}
where $F_p^{(A)}$ ($F_p^{(D)}$) is the Purcell enhancement for air-like (diamond-like) modes and $w_0$ is the $1/e^2$ intensity radius of the cavity waist. Note that in the limit $n_d \rightarrow 1$, these match what one would obtain from the standard Purcell formula $(3\lambda^3/4\pi^2)(Q/V)$ with a mode volume of $V = (\pi/4) w_0^2 L$~\cite{Gerard2003} and $Q = 2L\mathcal{F}/\lambda$. 

The two types of modes have different Purcell factors because they have different vacuum electric field maxima in the diamond and different cavity linewidths in frequency. The variation in $\Delta \nu$ has contributions from the reduced reflectivity of the flat mirror (due to the diamond layer) as well as the relative round-trip times of the diamond and air half-cavities.  
Such effects are similar to finesse oscillations observed in optomechanical systems~\cite{Wilson2009}. Remarkably, in the high finesse limit of the lossless 1D model, the length dependence of the vacuum electric field maximum in diamond precisely cancels the length dependence of $\Delta\nu$, yielding the simple expressions above.

For lossless systems, $\mathcal{F}$ matches the peak value of $\mathcal{F}_L$, and one might be tempted to use  Eqs.~\ref{purcell1}-\ref{purcell2} with our observed peak finesse and cavity geometry to determine the Purcell enhancement of our device. Such a calculation (using $\mathcal{F}=17,000$, $w_0$=2.2 \um, $\lambda =637$ nm) would predict $F_p^{(A)} \approx20$ and $F_p^{(D)} \approx 33$. However, adding in loss does not simply reduce $\mathcal{F}$: the location of the loss (in diamond or air) will affect the modes differently, and in general we find that using Eqs.~\ref{purcell1}-\ref{purcell2} with $\mathcal{F} = \mathrm{Max}(\mathcal{F}_L)$ overestimates the best Purcell enhancement for realistic systems where loss is associated with the diamond. 

Figure~\ref{fig3}c shows the Purcell factor calculated using the numerical model with the different loss mechanisms described in the previous section. While mirror absorption produces similar results to the predictions of Eqs.~\ref{purcell1}-\ref{purcell2}, qualitatively distinct behavior appears from the surface losses that likely limit our system. In particular, we predict a maximum Purcell enhancement of approximately 20 for our current device geometry. However, our analysis also suggests that significant improvements can be obtained. For example, if surface losses can be limited to the observed roughness after ArCl$_2$ etching ($< 0.2$ nm-rms), and higher reflectivity mirror coatings are used, a cavity finesse of $50,000$ can be maintained even with an antinode at the air-diamond interface. Using a $30~\mu$m radius-of-curvature mirror (attainable in our laser ablation setup), a $\langle 111\rangle$-oriented $5~\mu$m thick membrane, and a cavity length of $10~\mu$m, a maximum Purcell factor of around 200 could be reached. Such a cavity would also couple efficiently to the fiber mode ($>90\%$ with perfect alignment and low-loss mirrors~\cite{Joyce1984}) and have a linewidth $\sim$300 MHz, which is large enough to accommodate minor spectral diffusion but small enough to resolve the excited-state structure of the NV center.

\section{Finesse Changes with Cavity Length}

Beyond the absorption and scattering processes considered above, a thick diamond membrane could also induce an additional, potentially important source of loss: mixing between transverse modes of the cavity. 
Our numerical model has assumed that the air-diamond interface follows the spherical wavefront of the cavity mode, allowing description of the cavity eigenstates in terms of two Gaussian beams in the diamond and air regions. The real planar interface, however, deviates from this requirement, and can thereby couple the TEM$_{00}$ mode into higher-order Hermite-Gaussian modes.  Because high-order transverse modes have a larger spatial extent, this mechanism could induce additional losses caused by clipping at the small fiber mirror. This type of loss would behave differently than those discussed previously because it would depend on the length of the cavity, with greater losses expected when higher-order modes approach degeneracy with the fundamental. 


To estimate such losses, we apply non-degenerate perturbation theory (see Appendix \ref{app_perturb}) to calculate the first-order eigenstates of the membrane-in-cavity.  The fraction of those eigenstates clipped at the fiber mirror can then be calculated to determine the loss per round trip. We begin with zero-order modes from 
 the three-dimensional model discussed in Appendix \ref{3d_model}. We then treat the volume of diamond between the curved, wavefront-matching surface and the true flat interface as the perturbative volume. The first order correction to the eigenstate is given by~\cite{Sankey2014}: 
\begin{equation}
\psi^{1} = \kappa_{00}\sum_{m\neq 00} \frac{\iiint \psi_m(\mathbf{r})\, (n_d^2 -1) \,\phi_{00}(\mathbf{r})\,d\mathbf{r}^3}{\kappa_{m}-\kappa_{00}}\psi_m \label{first_corr}
\end{equation}
where the eigenstate is $\psi \approx \phi_{00}+\psi^1$, $\phi_{00}$ is a TEM$_{00}$ zero-order cavity mode derived from our model, $\psi_m$ is the $m^{th}$ zero-order mode (including all longitudinal and transverse mode numbers), $\kappa_m =\big( \frac{\omega_m}{c}\big)^2$ contains the eigenfrequency $\omega_m$ for the $m^{th}$ mode,  
and the integral is taken over the volume of the perturbation. 
The overlap integral couples only even order transverse modes, and falls off quickly with transverse mode number. In practice, we have included transverse mode numbers whose sum is $\leq 6$. 
Because the denominator grows quickly as the mode frequencies diverge, we consider corrections only from the two longitudinal modes closest in frequency to $\omega_{00}$. 

\begin{figure}[h!]
\includegraphics{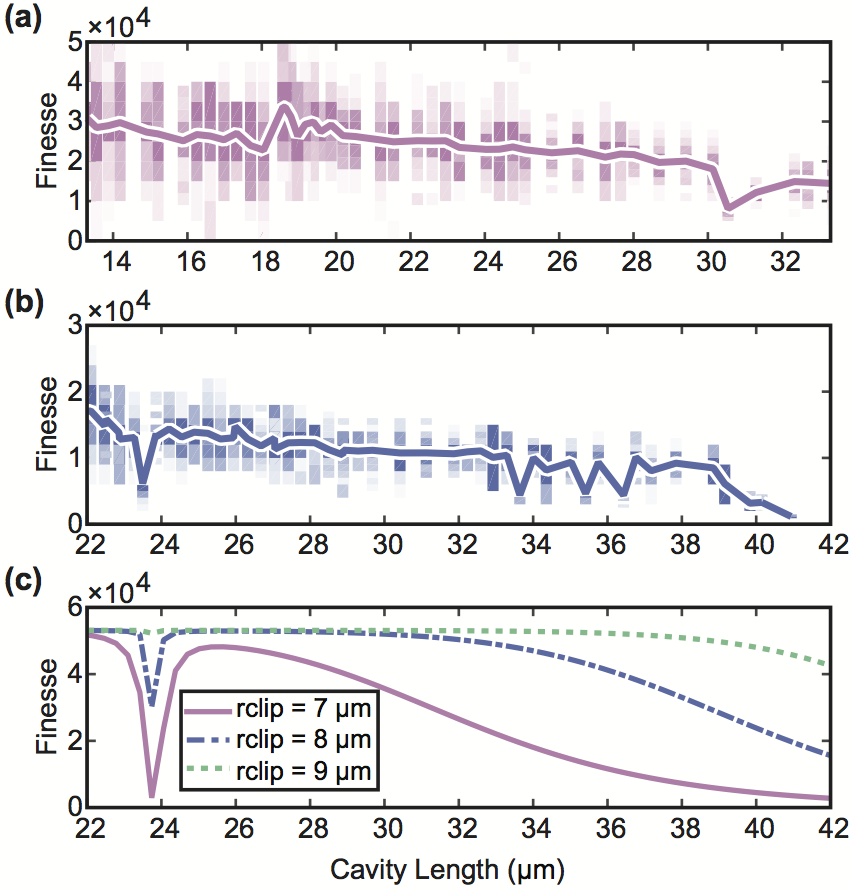}
\caption{(Color online) (a) Finesse as a function of length measured for a bare cavity. In (a-b), binned raw finesse data is shown through opacity, while the mean finesse as a function of length is indicated by the solid line. (b) Finesse as a function of length measured for a cavity containing a 10.5 \um~diamond membrane. The length axis is changed from (a) so that the beam radius on the fiber mirror takes on the same values over the range of collected data. (c) A simulation of finesse as a function of length including perturbative coupling to higher order modes evaluated for different mirror clipping radii. \label{fig4}}
\end{figure}

To examine the importance of these perturbative couplings experimentally, we measured finesse as a function of cavity length with and without the membrane (see Fig.~\ref{fig4}a-b). The bare cavity finesse exhibits a decreasing slope as a function of cavity length, which arises from coupling to lossy higher order modes caused by the Gaussian shape of the fiber dimple \cite{Benedikter2015_NJP,Kleckner2010_PRA}. 
We offset the x-axes of the bare cavity and membrane-in-cavity so that the beam radii on the fiber mirror would match, varying from 2.5 $\mu$m to 3.6 $\mu$m over the length range presented in Fig.~\ref{fig4}b-c. In addition to the decreasing slope seen for the bare cavity, we measured intermediate drops in finesse at specific cavity lengths. Since we use the same fiber mirror in both data sets, these dips must be associated with the membrane, and 
could be caused by coupling to lossy higher-order transverse modes.  For comparison, we simulated the cavity finesse (ignoring all other loss processes) using the first order correction to the electric field wavefunction (Eq.~\ref{first_corr}) for different clipping radii on the fiber mirror, outside of which all light is assumed to scatter out of the cavity. The result is shown in Fig.~\ref{fig4}c, which exhibits qualitatively similar drops in finesse at certain resonant lengths. 

We lack the detailed surface profile data to accurately parameterize our membrane and fiber dimple topography, so the simulations cannot include the exact perturbations present in the measurement (for example, it likely also contains a wedge, which would couple TEM$_{00}$ and TEM$_{10}$ modes). Nevertheless, our calculations demonstrate that the finesse reductions observed at specific cavity lengths in Fig.~\ref{fig4}b could reasonably be caused by this mechanism. Perhaps more importantly, this data illustrates that perturbative losses are not a major impediment to working with planar membranes, even at relatively large thicknesses $>10$ \um~and over a range of cavity lengths. Furthermore, as the diamond thickness diminishes, the perturbative coupling drops, indicating that it should be a negligible effect for few-micron-thick membranes.\\

\section{Conclusion}

We have shown that high finesse $\sim 17,000$ can be maintained in a Fabry-Perot microcavity even with incorporation of a thick diamond membrane. The membrane modifies the cavity modes, leading to variations in linewidth for different membrane thicknesses or different resonant frequencies. Our simulations indicate that surface losses dominate, producing qualitatively different behavior from bulk absorption. Membrane-induced coupling to higher-order transverse modes, conversely, does not greatly impact device performance. We anticipate that, despite the large surface losses, the current cavity will allow in the range of a 20-fold Purcell enhancement for diamond-based emitters, which in the case of the NV center would direct more than a third of its emission into the zero phonon line. Furthermore, device performance in this case is limited by surface roughness or contamination that is well above the currently attainable limits~\cite{Ovar2012_APL, Tao2013_AM}. 

When cooled to cryogenic temperatures and locked to the NV resonance frequency, such devices could significantly enhance the efficiency of photon-mediated entanglement between distant defects~\cite{Bernien2013}. Moreover, the cavity linewidth is below the typical spacing between spin-resolved resonant optical transitions in the NV center, enabling exploration of spin-dependent cavity effects. With improved diamond surface treatment and higher reflectivity mirrors, finesse $\sim 10^5$ should be possible~\cite{Hunger2010_NJP}, and shorter cavities with smaller radius-of-curvature mirrors could enhance the cavity cooperativity by another order of magnitude~\cite{Johnson2015}; even with current fabrication capabilities, Purcell enhancements in the range of 200 appear within reach. Ultimately, this highly-tunable open-cavity geometry could offer a route towards an efficient or even deterministic interface between single photons and solid-state spins.


\begin{acknowledgments}
We acknowledge support from NSERC, FRQNT, Canadian Foundation for Innovation, Canada Research Chairs, and INTRIQ. 
\end{acknowledgments}

\appendix
\section{Estimation of Bare Cavity Parameters}
\label{appendix_coupling}

\begin{figure}[h!]
\includegraphics{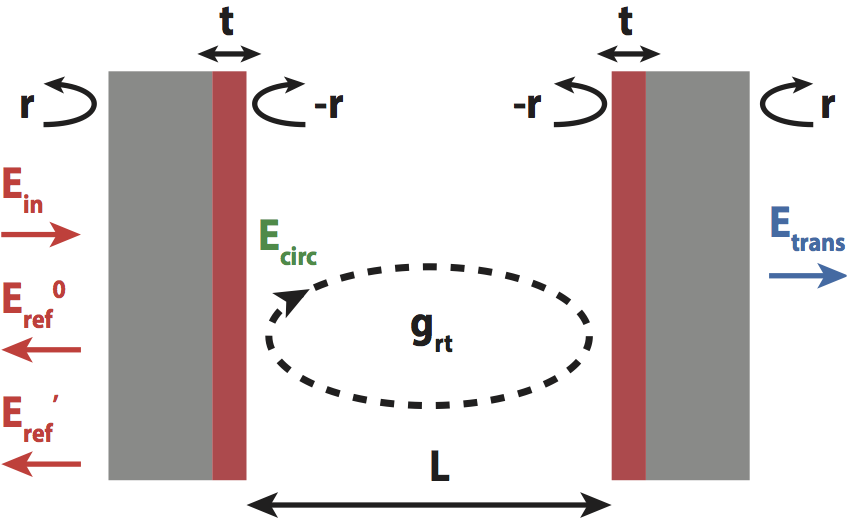}
\caption{\label{coupling_coeff} (Color online) The one-dimensional model used to estimate bare cavity parameters. $L$ is the cavity length; $r$ and $t$ are the mirror amplitude reflection and transmission coefficients, $g_{rt}$ is the round-trip gain of the cavity, and $E_{\mathrm{in}}$, $E_{\mathrm{ref}}^0$, $E_{\mathrm{ref}}'$, $E_{\mathrm{trans}}$, and $E_{\mathrm{circ}}$ are the electric fields at the input, reflected into the fiber, reflected into other modes, transmitted, and circulating in the cavity, respectively.}
\end{figure}

The following section describes a method for extracting some parameters of the bare fiber cavity system given measurable quantities and a one dimensional model shown in Fig.~\ref{coupling_coeff}.  These calculations permit us to understand the asymmetric lineshapes observed for both the bare and membrane-in-cavity systems.

In this model, the mirrors have real amplitude transmission and reflection coefficients $t$ and $r$, and we account for a loss per round trip in the cavity of $1-e^{-2\alpha}\approx 2\alpha$. Assuming that light is launched into the cavity through the fiber, $E_{\mathrm{in}}$ is the incident electric field, $E_{\mathrm{ref}}^0$ is the reflected field that is coupled back into the fiber core, $E_{\mathrm{ref}}'$ is the reflected field that is not coupled into the fiber core, and $E_{\mathrm{trans}}$ is the field coupled to the (free space) transmitted mode. The field circulating within the cavity is represented by $E_{circ}$, which is defined just to the right of the left-hand mirror;  the change in amplitude and phase incurred in one roundtrip is represented by the rountrip gain: $g_{rt}=e^{-2\alpha}e^{-\frac{2iL\omega}{c}}$. We consider imperfect cavity coupling, where $\epsilon_1$ is the overlap between the fiber and cavity modes, and $\epsilon_2$ is the coupling coefficient between the cavity and transmitted modes (for our analysis we set $\epsilon_2 \approx 1$). If the ablation spot is not perfectly centered on the fiber core, the promptly reflected light that is coupled back into the fiber mode can be described by a complex coupling coefficient $\eta$, which has a magnitude less than unity as well as a nonzero phase for imperfect alignment. The relevant relationships between these parameters are given by:
\begin{align}
 &E_{\mathrm{ref}}^0=-e^{-2\alpha-\frac{2iL\omega}{c}}rt\epsilon_1E_{circ}+r\eta E_{in}\\
 &E_{\mathrm{circ}}=e^{-2\alpha-\frac{2iL\omega}{c}} r^{2} E_{circ}+t\epsilon_{1} E_{in}\\
 &E_{\mathrm{trans}}=e^{-\alpha-\frac{iL\omega}{c}}t\epsilon_{2}E_{circ}
\end{align}
Solving for the transmitted and reflected powers ($|E_{trans}|^2$ and $|E_{ref}^0|^2$), normalized to the input power ($|E_{in}|^2$) yields the power reflected ($P_r$) and transmitted ($P_t$):
\begin{align}
&P_r = \frac{r^2\Big|\big(t^2\,\epsilon_1^2+\big(-e^{2\alpha+\frac{2iL\omega}{c}}+r^2\big)\eta\big)\Big|^2}{e^{4\alpha}+r^4-2\,e^{2a}\,r^2\,\text{cos}\big[\frac{2L\omega}{c}\big]}\\
&P_t =\frac{e^{2\alpha}\,t^4\,\epsilon_1^2\epsilon_2^2}{e^{4\alpha}+r^4-2\,e^{2\alpha}\,r^2\,\text{cos}\big[\frac{2L\omega}{c}\big]}.
\end{align}
We set $\eta=a+ib$ and expand the cosine terms to second order in $\Delta L$, where $L=m \lambda/2 + \Delta L$ and $m$ is an integer, resulting in power lineshapes of the form:
\begin{align}
&P_r = \frac{(a_1+a_2\,\Delta L)}{\pi}\frac{\big(\frac{\delta L}{2}\big)^2}{\big(\frac{\delta L}{2}\big)^2+\Delta L^2}+y_0\\
&P_t =  \frac{a_3}{\pi}\frac{\big(\frac{\delta L}{2}\big)^2}{\big(\frac{\delta L}{2}\big)^2+\Delta L^2}
\end{align}
where $\delta L$ is the FWHM cavity linewidth measured in length, and:
\begin{align}
&y_0 = \big(a^2+b^2\big)r^2+at^2\epsilon_1^2\\
&a1 = \pi t^2 \epsilon_1^2 \frac{a\big(r^4-e^{4\alpha}\big)+r^2t^2\epsilon_1^2}{\big(e^{2\alpha}-r^2\big)^2}\\
&a2 = \frac{4\pi b \,e^{2\alpha}\,r^2t^2\epsilon_1^2\omega}{c\big(e^{2\alpha}-r^2\big)^2}\\
&a3 = \pi \frac{e^{2\alpha}\, t^4\epsilon_1^2\epsilon_2^2}{(e^{2\alpha}-r^2\big)^2}
\end{align}
Note that this produces a Fano lineshape in reflection. The parameters $\{y_0, a_1, a_2, a_3\}$ can be extracted from our data by fitting the transmission and reflection curves and calibrating the input power. In addition, we use measurements of the finesse $\mathcal{F}$ and the following relationships to fully constrain the cavity parameters:
\begin{align}
&\mathcal{F} = \frac{\pi}{\alpha+t^2}\\
&t^2+r^2+\alpha=1,
\end{align}
as well as the known laser frequency $\omega$ and $\epsilon_2 = 1$. With these expressions, one can solve for all of the cavity parameters of interest.  For example, using data acquired in a bare cavity of length $12.2 \pm 0.3~\mu$m  
we obtain:

\begin{align*}
t&=(8.8\pm0.2)*10^{-3}\\
r&=-0.999957\pm0.000001\\
a&=0.61\pm0.02\\
b&=0.14\pm0.04\\
\epsilon_1&=0.69\pm0.03\\
\alpha&=8\pm1 \text{ ppm}
\end{align*}
This yields a power transmittance of $T=78\pm3\text{ ppm}$, which agrees with the quoted coating value of $T=70\pm10\text{ ppm}$. The combined absorption and scattering losses were quoted to be $<24\text{ ppm}$, which also agrees with the derived $\alpha$ value.

Note: While in the final stages of preparing this manuscript, we became aware of detailed theoretical and experimental exploration of the origin of asymmetric lineshapes associated with misalignment of fiber cavities~\cite{recentpaper}.

\section{Calculating the Effective Radius of Curvature}

\label{app_rcurvature} 

The Gaussian-shaped ablation dimple can be approximated by a parabola near the center, which has a well defined radius of curvature. This radius is appropriate for cavity modes with small beam diameters on the mirror. As the cavity length is increased, the mode diameter grows and the approximation breaks down. In this regime, it is more accurate to estimate the effective mirror radius from the spacing of the higher order TEM modes. If $\Delta\nu_{trans}$ is the difference in frequency between adjacent transverse modes with the same longitudinal mode (e.g. TEM$_{m,n}$ and TEM$_{m, n+1}$), the effective radius of curvature is
\begin{align}
&R = L\bigg(1-\text{cos}^2\bigg(\, \frac{\Delta \nu_{trans}}{\nu_{FSR}}\pi\bigg)\bigg)^{-1},
\end{align}
where $L$ is the length of the cavity and $\nu_{FSR} = c/2L$ is the free spectral range. Using this equation to analyze the  TEM$_{00}$ and TEM$_{01}$/TEM$_{10}$ modes in the white-light spectrum measured for the bare cavity, we estimate an effective radius of curvature of $61.0\,\pm1.4\ $ \um~for a bare cavity length of 13.3 \um~and a beam radius of $2.6$ \um~on the fiber mirror. The same beam diameter would correspond to a cavity length of $22\ $ \um~for a cavity containing a $10.5 $ \um~diamond membrane, as the beam diverges less in the higher refractive index medium. Deviation from the radius of curvature extracted from a parabolic fit to our interferometry measurement ($R \approx50\pm1~\mu$m) arises because the finite diameter mode samples a range of curvatures within the Gaussian dimple. 

\section{Numerical Cavity Model}
\label{3d_model}
\begin{figure}[h!]
\includegraphics[scale = 0.7]{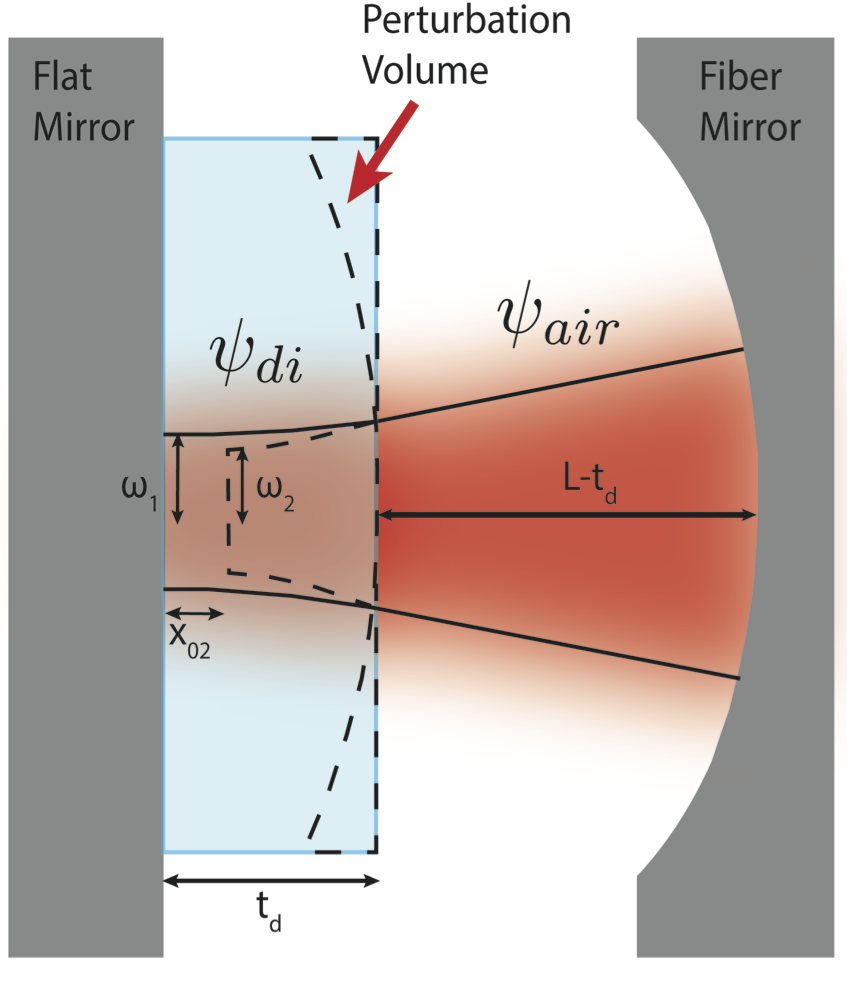}
\caption{(Color online) A diagram illustrating the $1/e^2$ intensity radius of the zero-order Gaussian modes in the diamond and air (solid lines). The mode in diamond has a waist with radius $w_1$ at the flat mirror, while the mode in air has a waist with radius $w_2$ a distance $x_{02}$ from the mirror flat. The perturbation volume considered in Eq.~\ref{first_corr} (dashed lines) is the difference between the presumed diamond interface lying along the mode wavefront and the planar diamond surface.\label{cavitymode}}
\end{figure}

We have developed a numerical model to calculate the fundamental Gaussian cavity mode for a half symmetric cavity containing a diamond membrane bonded to the flat mirror. The model first solves for the Gaussian beam parameters (waist radius and position) in both the air and diamond sections assuming a curved diamond surface lying along the mode wavefront (see Fig.~\ref{cavitymode}). 
As boundary conditions, we require that the beam diameters and radii of curvature are equal at the air-diamond interface to ensure electric field continuity. In addition, the radius of curvature in air should match the ablation radius of curvature at the fiber mirror, while the mode waist in diamond should lie at the flat mirror. With these four requirements, once can solve for the required radius of curvature for the air-diamond interface, the beam waists $w_1$ and $w_2$ corresponding to the modes in diamond and air, and the effective waist position $x_{02}$ for the air mode. 

\begin{figure}[h!]
\includegraphics{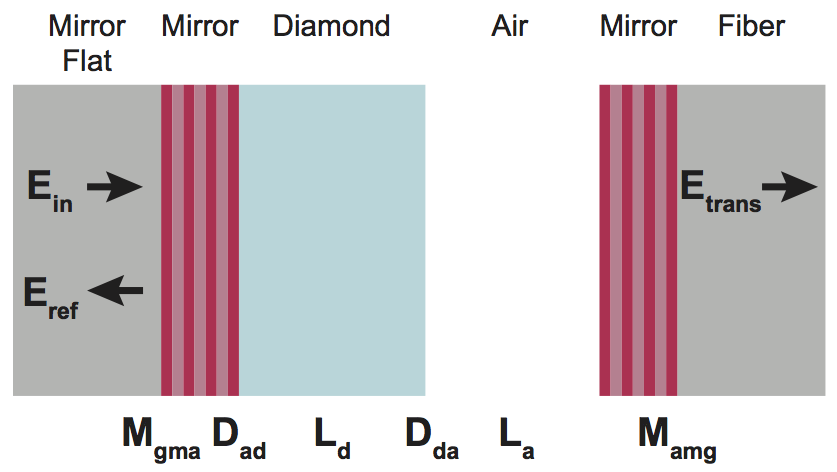}
\caption{(Color online) A diagram indicating the relevant transfer matrices used to calculate the resonant cavity frequencies and lengths. \label{fig_transfermatrix}}
\end{figure}

Once the two Gaussian modes in the air and diamond regions are known, one can solve for the resonant frequencies and lengths of the cavity using transfer matrix theory applied to the left and right traveling electric fields within the cavity structure (see Fig.~\ref{fig_transfermatrix}).

Here, we model the full geometry of the flat mirror as a one-dimensional 29 layer dielectric stack where the right-moving electric field travels from glass into air; the transfer matrix describing this process is $M_{gma}$. Conversely, the fiber mirror is modeled as the inverse matrix ($M_{amg} = M_{gma}^{-1}$). Since the mirrors are defined for air termination, we need to include an additional matrix to model the air-diamond interface at the flat mirror ($D_{ad}$). As mentioned above, we assume that the second air-diamond interface ($D_{da}$) follows the curvature of the wavefronts at that position (dashed line in Fig.~\ref{cavitymode}). The propagation matrices in the air and diamond include Guoy phase, and are given by:
\begin{align}
&L_{diamond} = \begin{bmatrix}
    e^{-\frac{2\pi n_d}{\lambda}t_d+i \phi_1(t_d)}       &  0 \nonumber  \\
   0    &  e^{\frac{2\pi n_d}{\lambda}t_d-i \phi_1(t_d)}  \nonumber \\
\end{bmatrix}\\ \nonumber \\ 
&L_{air} = \begin{bmatrix}
    e^{-\frac{2\pi}{\lambda}(L-t_d)+i \phi_2(L-x_{02})}       &  0  \nonumber \\
   0    &   e^{\frac{2\pi}{\lambda}(L-t_d)-i \phi_2(L-x_{02})} \nonumber \\
\end{bmatrix}
\end{align}
where $\phi_i(x) = \arctan{\left(x\lambda/\pi w_i^2\right)}$ is the Guoy phase, $t_d$ is the diamond thickness,  $L$ is the cavity length, $w_1$ and $w_2$ are the $1/e^2$ radii at the diamond and air waists respectively, and $x_{02}$ is the effective beam waist position for the mode in air (see Fig.~\ref{cavitymode}). The full transfer matrix for the cavity is:
\begin{align}
\begin{bmatrix}
    &E_{trans}  \\
   &0  
\end{bmatrix}  =
S
\begin{bmatrix}
    &E_{in}  \\
   &E_{ref} 
\end{bmatrix}
\end{align}
where:
\begin{align}
S = M_{amg}\,L_a\,D_{da}\,L_d\,D_{ad}\,M_{gma}.
\end{align}
Using the transmission curves calculated with this model, the linewidth in frequency and length can be determined.  The field within the diamond and air regions can be found by evaluating subsets of the transfer matrices to find the amplitudes in the air and diamond, which are then multiplied by appropriate Gaussian modes.

When adding loss to this model, the dielectric indices used in the mirror stack and diamond were given small complex components; in addition the interface matrices $D_{da}$ and $D_{ad}$ were modified according to Eqs.~\ref{rij}-\ref{tij}.




\section{Nondegenerate Perturbation Theory}
\label{app_perturb}
The Hermite-Gaussian family of modes represent the eigenstates of a spherical resonator, satisfying the Helmholtz equation \cite{saleh1991fundamentals}. 
 Introducing some small volume of material with a different refractive index can break the cavity symmetry, leading to a new set of cavity eigenstates that can be expressed as a linear combination of the unperturbed cavity modes~\cite{Sankey2014}. 
 
 In our case, the zero-order modes correspond to solutions in the presence of a membrane whose interface is curved to follow a wavefront (see Fig.~\ref{cavitymode}). They are defined by
\begin{equation}
\left(\nabla^2+\kappa_i n_0^2(\mathbf{r})\right)\psi_i^0(\mathbf{r})=0,
\end{equation}
where $\psi_i^0$ are the zero-order modes of the system,  $\kappa_i=(\frac{\omega_i}{c})^2$ contains the corresponding eigenfrequencies $\omega_i$, and $n_0^2(\mathbf{r})$ is the index of refraction inside the cavity assuming an air-diamond interface following the mode wavefront. These zero-order eigenfunctions may be found exactly as a (real-valued) Hermite-Gaussian family of modes with different parameters in the air and diamond regions (see Appendix~\ref{3d_model}).  The orthonormalization condition is
\beq
\iiint\psi_m^0(\mathbf{r}) n_0^2(\mathbf{r}) \psi_n^0(\mathbf{r}) d^3\mathbf{r} = \delta_{mn},
\eeq
where $\delta_{mn}$ is the Kronecker delta, the integral is taken over the cavity volume, and subscripts $m$ and $n$ encode all transverse and longitudinal mode indices.

We wish to calculate the perturbative effect of the membrane planarity, which is equivalent to introducing a small piece of dieletric representing the difference between the curved surface and a flat one. We are interested in a fundamental transverse mode (which is non-degenerate), for which the exact eigenstate $\psi$ satisfies
\begin{equation}
\left(\nabla^2+\kappa\left(n_0^2(\mathbf{r})+\lambda \mathcal{V}(\mathbf{r}) \right)\right)\psi=0,
\label{perturb}
\end{equation}
where $\mathcal{V}(\mathbf{r}) = n_d^2-1$ inside the perturbation volume (and zero outside), $\lambda$ is some small number, and $\kappa$ corresponds to the new eigenfrequency. We can express both $\psi$ and $\kappa$ as a power series in $\lambda$:
\begin{align}
\psi &= \psi_0^0 + \sum_{n=1}^{\infty}\lambda^n \psi^n\\
\kappa &= \kappa_0 + \sum_{n=1}^{\infty}\lambda^n \Delta_n
\end{align}
where $\psi_0^0$ and $\kappa_0$ correspond to the non-degenerate fundamental mode eigenstate of the unperturbed system, and $\psi^n$, $\Delta_n$ are the $n^{th}$ order corrections. Considering only the terms of Eq.~\ref{perturb} to order $\lambda$, one finds
\begin{equation}
( \nabla^{2} + n_0^2(\mathbf{r})\kappa_{0})\psi^{1} = -(\Delta_1n_0^2(\mathbf{r}) + \kappa_{0}\mathcal{V}(\mathbf{r}) )\psi_{0}^0,
\end{equation}
and one can thereby derive the first order correction to the eigenstate:
\begin{equation}
\psi^1 = \kappa_{0} \sum_{m\neq 0} \frac{\iiint \psi_m^0(\mathbf{r}) \, \mathcal{V}(\mathbf{r}) \, \psi_0^0(\mathbf{r}) \, d\mathbf{r}^3}{\kappa_m - \kappa_{0}}\, \psi_m^0
\end{equation}
where $\psi_m^0$ is the $m$th order mode of the unperturbed system, and we have set $\lambda = 1$. Here $m$ labels all longitudinal and transverse modes to which the zero-order Gaussian mode $\psi_0^0$ can be coupled by the perturbation.


\bibliography{diamond_mem_cavity_ArXiv}

\end{document}